\documentclass[letterpaper, 10 pt, conference]{ieeeconf}  

\IEEEoverridecommandlockouts                              

\usepackage{graphicx}
\usepackage[utf8]{inputenc}
\usepackage{ntheorem}
\usepackage{amsmath}
\usepackage{amssymb}
\usepackage{xcolor}	
\usepackage{cite}
\usepackage{dsfont}
\usepackage{amsfonts}
\usepackage{subfig}
\usepackage{float}
\usepackage{url}
\usepackage{cite}

\newtheorem*{definition}{Definition}

\title{\LARGE \bf
Ratiometric control for differentiation of cell populations endowed with synthetic toggle switches
}

\author{Davide Salzano$^{1}$, Davide Fiore$^{1}$, Mario di Bernardo$^{1,2}$
\thanks{$^{1}$Davide Salzano, Davide Fiore and Mario di Bernardo are with the Department of Electrical Engineering and Information Technology, University of Naples Federico II, Via Claudio 21, 80125 Naples, Italy.
        {\tt\small davidesalzano94@gmail.com, davide.fiore@unina.it}}%
\thanks{$^{2}$Mario di Bernardo is also with the Department of Engineering Mathematics, University of Bristol, University Walk, BS8 1TR Bristol, U.K.
        {\tt\small mario.dibernardo@unina.it}}%
}

\begin{document}

\maketitle
\thispagestyle{empty}
\pagestyle{empty}

\begin{abstract}
We consider the problem of regulating by means of external control inputs the ratio of two cell populations. Specifically, we assume that these two cellular populations are composed of cells belonging to the same strain which embeds some bistable memory mechanism, e.g. a genetic toggle switch, allowing them to switch role from one population to another in response to some inputs. We present three control strategies to regulate the populations' ratio to arbitrary desired values which take also into account realistic physical and technological constraints occurring in experimental microfluidic platforms. The designed controllers are then validated in-silico using stochastic agent-based simulations.
\end{abstract}


\section{Introduction}
The aim of Synthetic Biology is to engineer biomolecular systems to achieve new useful functionalities \cite{del2018future}. Potential applications range from designing bacteria that can produce biofuels or sense and degrade pollutant in the environment (like hydrocarbons and plastic), to immune cells that can track and kill cancer cells, or that can release drugs at specific points and conditions to avoid side effects (see \cite{del2018future} for references).
This is possible by designing genetic circuits with programmed functionalities and embedding them into living cells. However, most of the engineered genetic circuits have been designed to work at single-cell level. As a consequence, their functional complexity is limited by inherent factors such as excessive metabolic burden on the cell, competition of limited resources, and incompatible chemical reactions.

A promising approach to overcome these issues is to engineer synthetic microbial consortia in which the effort is divided and assigned to different subpopulations of cells to achieve more sophisticated functionalities \cite{bittihn2018rational}.  Recent cooperative consortia designs include a predator-prey system \cite{balagadde2008synthetic}, an emergent oscillator \cite{chen2015emergent}, a toggle-switch implemented across two species \cite{sadeghpour2017bistability}, and a multicellular feedback control scheme where the control functions are split between two species \cite{fiore2016silico}.
Unfortunately, the correct functioning of a multicellular consortium requires cocultivating and maintaining multiple cell populations. As different cells in the consortium embed specific sets of genetic circuits, they also present different growth rates due to uneven metabolic burdens and might show additional undesired dynamics, such as oscillations \cite{sadeghpour2017bistability}. Therefore, when different strains are mixed together, it is essential to maintain their stable coexistence by controlling their relative population numbers (i.e. their ratio). This is usually achieved by encoding in the synthetic design some dynamic equilibrium between the two populations, e.g. \cite{igem},\cite{ren2017population}.
However, if one of the two populations eventually dies out, these solutions can either lead to uncontrolled growth of one population or to the extinction of both. Moreover, the steady-state value of the populations ratio is hard-coded into the genes without any possibility of being changed online.

In this paper we present an alternative approach to control the populations' ratio in mono-strain consortia by means of external control inputs. Specifically, we consider the case in which there exists a bistable memory mechanism inside the cells, such as the genetic toggle switch circuit \cite{Gar}, whose current internal state defines which of the two possible roles, or “working-condition”, the cell is playing in the consortium. We assume that, by changing the concentrations of some inducer molecules in the growth medium, it is possible to make cells switch their role and hence keep the populations ratio to a desired value. 
Albeit requiring a possibly more complex design with respect to other multi-strain scenarios, this approach has the advantage of being  intrinsically robust to extinction events that could undermine the operation of the entire consortium. Also, it allows the ratio of the populations to be changed online, in real-time, if needed.

The crucial problem we address in this paper is to design feedback control strategies able to steer the inducer molecules inputs to achieve and maintain a desired cell ratio. We define this problem as \emph{``ratiometric"} control of cell populations.  

We propose and test three different external control strategies to regulate the populations ratio to any value, namely a Bang-Bang controller, a PI controller and a model predictive controller (MPC). 
The control laws are designed taking into account realistic physical and technological implementation constraints that are present in microfluidic-based experimental platforms.  
Finally, the proposed controllers are validated \emph{in-silico} using realistic stochastic agent-based simulations in BSim \cite{BSim} that appropriately model also spatial and diffusion effects in the microfluidic device and cell growth.
%
%
%
\section{The ratiometric control problem}
We want to design some feedback control strategy such that, by acting on some input signals \emph{common} to every cells, the ratio between the two populations is asymptotically regulated to some desired value.
%
%
%
%
\begin{figure}[t]
	\centering
	\includegraphics[width=0.8\linewidth]{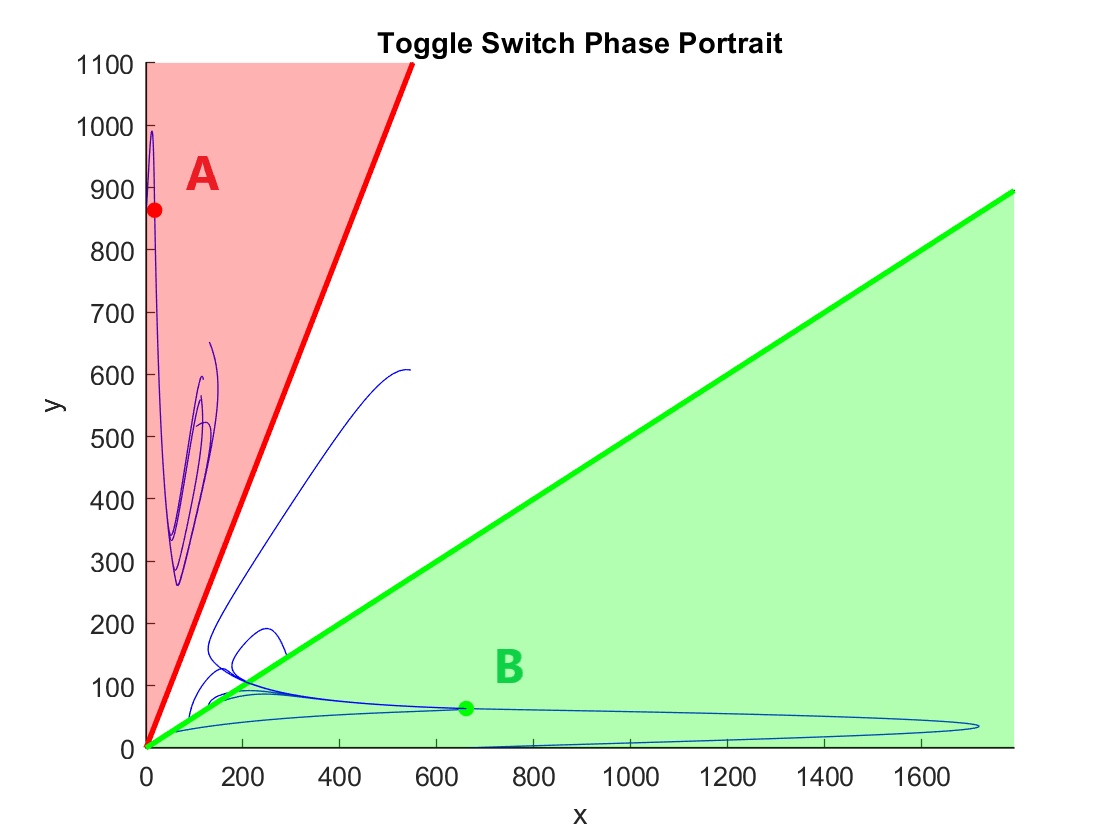}
	\caption{Regions of the state space $(LacI^i, TetR^i)$ such that cells belong to set $\mathcal{A}_t$ (red color) or $\mathcal{B}_t$ (green color). The positions of the stable equilibrium points \textbf{A} and \textbf{B} are reported in the phase plane with several examples of solutions starting from different initial conditions.}
	\label{fig:phasePortraitLugagne}
\end{figure}
\subsection{Population model} \label{sec:Model}
We assume that the bistable memory required by the cells to guarantee their the correct operations in the consortium is realized by means of an inducible genetic toggle switch \cite{Gar}.
This genetic regulatory network consists of two repressor proteins, LacI and TetR, both repressing each other’s promoter, so that only one protein is fully expressed at any time. From a modelling viewpoint, the genetic toggle switch is a bistable dynamical system, possessing two stable equilibria, \textbf{A} and \textbf{B}, each associated to a fully expressed protein, and a saddle equilibrium point, whose stable manifold is the boundary separating the regions of attraction of the other two.
Thus, given an initial condition, its solutions will converge to one of the two stable equilibria.
The expression level of the two repressing proteins can be flipped by changing the concentration of two inducer molecules, aTc and IPTG.
This causes the occurrence of two saddle-node bifurcations yielding the required reversible bistable memory function.
We use the inducible toggle switch model described in \cite{lugagne} and further analyzed in \cite{fiore2019analysis,guarino2018silico}. Namely, we assume the dynamics of the $i$-th cell in the consortium can be written as follows:

\vspace{-0.3cm}
\footnotesize
\begin{align}
\label{eq:LugagneOriginal}
& \frac{\mathrm{d}\, mRNA_\mathrm{LacI}^{i}}{\mathrm{dt}} = \kappa_\mathrm{L}^\mathrm{m0} + \kappa_\mathrm{L}^\mathrm{m} \, \Phi_\mathrm{T}(t)-\gamma_\mathrm{L}^\mathrm{m} \, {mRNA_\mathrm{LacI}}\\
& \frac{\mathrm{d}\, mRNA_\mathrm{TetR}^{i}}{\mathrm{dt}} = \kappa_\mathrm{T}^\mathrm{m0} + \kappa_\mathrm{T}^\mathrm{m} \, \Phi_\mathrm{L}(t) - \gamma_\mathrm{T}^\mathrm{m} \, {mRNA_\mathrm{TetR}}\\
& \frac{\mathrm{d}\, LacI^{i}}{\mathrm{dt}} = \kappa_\mathrm{L}^\mathrm{p} \, {mRNA_\mathrm{LacI}^{i}} - \gamma_\mathrm{L}^\mathrm{p} \, {LacI^{i}}\\
& \frac{\mathrm{d}\, TetR^{i}}{\mathrm{dt}} = \kappa_\mathrm{T}^\mathrm{p} \, {mRNA_\mathrm{TetR}^{i}} - \gamma_\mathrm{T}^\mathrm{p} \, {TetR^{i}}\\
& \frac{\mathrm{d}\, aTc^{i}}{\mathrm{dt}} = k_\mathrm{aTc} \, \left(u_\mathrm{a} - {aTc^{i}} \right)\\
\label{eq:LugagneOriginal_last}
& \frac{\mathrm{d}\, IPTG^{i}}{\mathrm{dt}} = k_\mathrm{IPTG} \, \left(u_\mathrm{p} - {IPTG^{i}} \right)
\end{align}
\normalsize
where the state variables denote concentrations of molecules inside the cell. The parameters $\kappa_\mathrm{L/T}^\mathrm{m0}$, $\kappa_\mathrm{L/T}^\mathrm{m}$, $\kappa_\mathrm{L/T}^\mathrm{p}$, $\gamma_\mathrm{L/T}^\mathrm{m}$, $\gamma_\mathrm{L/T}^\mathrm{p}$, $k_\mathrm{aTc/IPTG}$ are leakage transcription, transcription, translation, mRNA degradation and protein degradation rates, and diffusion rates of the inducers across the cell membrane, respectively. 
The variables $u_{\mathrm{a}}$ and $u_{\mathrm{p}}$ denote the concentrations of the inducer molecules in the growth medium and they also represent the control inputs common to every cell in the populations.
Moreover, in the previous equations, the input effects are modelled by the following terms:
\begin{align*}
    & \Phi_\mathrm{T}(t) : = \frac{1}{1+\left(\frac{TetR^{i}}{\theta_{\mathrm{TetR}}}\cdot \frac{1}{1+\left(\frac{aTc^{i}}{\theta_{\mathrm{aTc}}}\right)^{\eta_{\mathrm{aTc}}}}\right)^{\eta_{\mathrm{TetR}}}}\\
    & \Phi_\mathrm{L}(t) : = \frac{1}{1+\left(\frac{{LacI^{i}}}{\theta_{\mathrm{LacI}}}\cdot \frac{1}{1+\left(\frac{{IPTG^{i}}}{\theta_{\mathrm{IPTG}}}\right)^{\eta_{\mathrm{IPTG}}}}\right)^{\eta_{\mathrm{LacI}}}}
\end{align*}
All parameter values of are provided in Table \ref{tab:par}  and are also the same used in \cite{lugagne}.
\begin{table}
	\centering
	\begin{tabular}{|c|c||c|c|}
		\hline 
		$ \kappa_\mathrm{L}^\mathrm{m0} $ & $ 3.045\cdot10^{-1} $ mRNA $\min^{-1}$& $ \gamma_\mathrm{L}^\mathrm{p} $ & $ 1.65\cdot10^{-2} $ $\min^{-1}$ \\
		\hline 
		$ \kappa_\mathrm{T}^\mathrm{m0} $ & $ 3.313\cdot10^{-1} $ mRNA $\min^{-1}$& $ \gamma_\mathrm{T}^\mathrm{p} $ & $ 1.65\cdot10^{-2} $ $\min^{-1}$  \\ 
		\hline 
		$ \kappa_\mathrm{L}^\mathrm{m} $ & $ 13.01 $  mRNA $\min^{-1}$&  $ \theta_{\mathrm{LacI}} $ & $ 124.9 $ \\ 
		\hline 
		$ \kappa_\mathrm{T}^\mathrm{m} $ & $ 5.055 $ mRNA $\min^{-1}$&   $ \eta_{\mathrm{LacI}} $ & $ 2.00 $ \\ 
		\hline 
		$ \kappa_\mathrm{L}^\mathrm{p} $ & $ 0.6606 $ a.u. $\text{mRNA}^{-1}$ $\min^{-1}$&   $ \theta_{\mathrm{TetR}} $ & $ 76.40 $ \\ 
	    \hline 
		$ \kappa_\mathrm{T}^\mathrm{p} $ & $ 0.5098 $ a.u. $\text{mRNA}^{-1}$ $\min^{-1}$&  $ \eta_{\mathrm{TetR}} $ & $ 2.152 $\\ 
		\hline 
		$ k_{\mathrm{aTc}} $ & $ 4\cdot 10^{-2} $ $\min^{-1}$ &    $ \theta_{\mathrm{aTc}} $ & $ 35.98 $ \\
		\hline
		$ k_{\mathrm{IPTG}} $ &  $ 4\cdot 10^{-2}  $ $\min^{-1}$ & $ \eta_{\mathrm{aTc}} $ & $ 2.00 $\\
		\hline
		$ \gamma_\mathrm{L}^\mathrm{m} $ & $ 1.386\cdot10^{-1} $ $\min^{-1}$ &  $ \theta_{\mathrm{IPTG}} $ & $ 2.926\cdot10^{-1} $\\
		\hline
		$ \gamma_\mathrm{T}^\mathrm{m} $ & $ 1.386\cdot10^{-1} $ $\min^{-1}$ &  $ \eta_{\mathrm{IPTG}} $ & $ 2.00 $\\
		\hline
	\end{tabular}  
	\caption{Value of the parameters of the cell population models (taken from \cite{lugagne}).}
	\label{tab:par}
\end{table}

The previous dynamical model is a deterministic description of the evolution of the molecule concentrations in the system and, therefore, it is only an approximation of the stochastic biochemical processes taking place inside the cells.
To obtain a more accurate description of the stochastic processes governing the dynamics, for validation we adopted the SDE-based algorithm described in \cite{lakatos2017stochastic} that provide a better approximation of the Chemical Master Equation \cite{CME,CLE} of the system.
Formally, we solved:
\begin{equation}
    \mathrm{d} \mathbf{x}(t)=S\cdot  a(\mathbf{x}(t))\cdot \mathrm{dt}+ S \cdot \mathrm{diag} \big( \sqrt{a(\mathbf{x}(t))} \big) \cdot \mathrm{d}\mathbf{w}
\end{equation}
where $\mathbf{x}(t)$ is the state of the process, $S$ is the stoichiometric matrix, $a(\mathbf{x}(t))$ is a vector containing the propensity functions associated to each reaction and $\mathbf{w}$ is a vector of independent standard Wiener processes. Both $S$ and $a(\mathbf{x})$ are the same used in \cite{lugagne}. \\
As shown later in the \emph{in-silico} validation of the control approaches, the heterogeneity in the response of the cells, provided in our case by the biochemical noise, is a fundamental ingredient to solve the ratiometric control problem.
%
%
%
\subsection{Problem Statement}
We denote by $\mathcal{N}_t$ the finite set of all cells in the consortium at time $t$, and with $N(t)=|\mathcal{N}_t|$ its cardinality. Note that the number of cells may vary in time as a consequence of cell births and deaths or of their accidental outflow from the microfluidic chamber in which they are hosted.
We define the following sets: $\mathcal{A}_t:=\{ i\in\mathcal{N}_t:\, TetR^i(t) > 2\, LacI^i(t) \}$, $\mathcal{B}_t:=\{ i\in\mathcal{N}_t:\, LacI^i(t) > 2\, TetR^i(t) \}$, and $\mathcal{C}_t:=\{ i\in\mathcal{N}_t:\, i\notin\mathcal{A}_t, i\notin\mathcal{B}_t \}$. 
We also denote with $n_\mathrm{A}(t)$ and $n_\mathrm{B}(t)$ the cardinalities of $\mathcal{A}_t$ and $\mathcal{B}_t$ at time $t$, respectively.
It is clear from Figure \ref{fig:phasePortraitLugagne} that these sets are disjoint and form a partition of $\mathcal{N}_t$ for all $t$.

Noticing the relative position in state space $(LacI^i, TetR^i)$ of the stable equilibria \textbf{A} and \textbf{B} of the toggle switch, we say that at time $t$ cell $i$ \emph{belongs} to the population \textbf{A} (population \textbf{B}), if $i\in\mathcal{A}_t$ ($i\in\mathcal{B}_t$, respectively).
Moreover, we define as $r_\mathrm{A}(t) = \frac{n_\mathrm{A}(t)}{N(t)}$ and $r_\mathrm{B}(t) = \frac{n_\mathrm{B}(t)}{N(t)}$ the \emph{ratio} of cells that belong to population \textbf{A} and population \textbf{B}, respectively. 

\begin{definition}
Given a consortium of cells whose dynamics is described by \eqref{eq:LugagneOriginal}-\eqref{eq:LugagneOriginal_last} and a desired ratio $r\in[0,1]$ of cells belonging to one population, for example \textbf{B}, we say that the control law $u(t)=\left[u_\mathrm{a}(t), \, u_\mathrm{p}(t)\right]^\top$ solves the \emph{ratiometric control problem} if, for some  small positive constant $\epsilon$,
\begin{equation}
\label{eq:control_aim}
\lim_{t\to \infty}|e_\mathrm{A}(t)| < \epsilon  \quad  \text{ and } \quad  \lim_{t\to \infty}|e_\mathrm{B}(t)|<\epsilon,
\end{equation}
where $e_\mathrm{B}(t) = r-r_\mathrm{B}(t)$ and $e_\mathrm{A}(t) = (1-r)-r_\mathrm{A}(t)$.
\end{definition}
%
%
%
%
\section{Proposed Control Strategies}
In this section we propose three control strategies to solve the ratiometric control problem. Specifically, we present a Bang-Bang controller, a PI controller and a model predictive controller (MPC).
All controllers are ad-hoc implementations that take explicitly into account the physical and technological constraints related to an experimental microfluidic platform.
Then, in Section \ref{sec:simulations} we validate the controllers {in-silico}.
In contrast to \cite{lugagne} where the objective of the controllers proposed was to regulate at an \emph{intermediate} level the expression of the genes of a \emph{single} toggle-switch, here the feedback loop is closed on the entire cell population and the cells are split in two groups, each fully expressing either of the two genes.
\subsection*{Experimental Constraints}
The experimental platform we consider as a reference is based on microfluidics as the one described in \cite{menolascina2014vivo,Perrino2015} which uses a fluorescence microscope to measure the current state of the cells.
We then have to take into account the following realistic constraints:
\begin{enumerate}
    \item
    the state of the cells cannot be measured more often than $5\,\mathrm{min}$ to avoid excessive phototoxicity;
    \item
    there is a time delay between $20$ and $40\,\mathrm{s}$ on the actuation of the control inputs due to the time that the flow of the chemical inducers takes to reach the chambers on the microfluidic chip where cells are hosted;
    \item
    the minimum time interval between two consecutive control inputs cannot be less than $15\,\mathrm{min}$ to limit excessive  osmotic stress on the cells;
    \item
    the maximum duration of any experiment cannot exceed $24$ hours, to avoid substantial cell mutations during the experiments.
\end{enumerate}
Moreover, the specific implementation of the microfluidic device also introduces constraints on the possible classes of input signals $u(t)=\left[u_\mathrm{a}(t), \, u_\mathrm{p}(t)\right]^\top$ that can be generated by the actuators.

We consider two possible implementations: 
\begin{enumerate}
    \item 
    a T-Junction, which limits $u_\mathrm{a}$ and $u_\mathrm{p}$ to be mutually exclusive and with fixed amplitudes, that is $u$ is either set to $[U_\mathrm{a},0]^\top$, which causes $e_\mathrm{B}$ to decrease and $e_\mathrm{A}$ to increase, or to $[0, U_\mathrm{p}]^\top$, which does the opposite; 
    \item 
    a Dial-A-Wave (DAW) system \cite{ferry2011microfluidics}, which constraints $u_\mathrm{a}$ and $u_\mathrm{p}$ to be in a convex combination. Namely, given $u_\mathrm{a} \in [0,U_a]$ we have 
    \begin{equation}
    \label{eq:DAW}
     u_\mathrm{p}=\left(1-\frac{u_\mathrm{a}}{U_\mathrm{a}}\right)U_\mathrm{p}
    \end{equation} 
\end{enumerate}
In the above equations, $U_\mathrm{a}\in\left[0,100\right]$ and $U_\mathrm{p}\in\left[0,1\right]$ are control amplitudes to be selected and denote the maximum concentrations possible of the inducers that are present in the reservoirs (These values are the same as those that were used \emph{in-vivo} in  \cite{lugagne}.).

Depending on which implementation is considered, only specific controllers are feasible. More precisely, for the T-Junction implementation only a Bang-Bang controller is considered, while for the Dial-A-Wave system we design a PI controller and an MPC.
%
%
%
%
\subsection{Bang-Bang Controller}
The Bang-Bang controller implemented via a T-junction consists of two mutually exclusive inputs with fixed amplitude which are applied to the system depending on the current value of the error signals $e_\mathrm{A}(t)$ and $e_\mathrm{B}(t)$.
Specifically, at any time $t$ the input is applied that causes the  $\max\{ |e_\mathrm{A}(t)|, |e_\mathrm{B}(t)| \}$ to decrease.
More formally, the control input $u(t)=\left[u_\mathrm{a}(t), \, u_\mathrm{p}(t)\right]^\top$ is chosen as
\begin{equation}
    u(t) = 
    \begin{cases}
        u_{1}, & |e_\mathrm{B}(t)|\geq |e_\mathrm{A}(t)| \\
        u_{2}, & |e_\mathrm{B}(t)| < |e_\mathrm{A}(t)|\\
    \end{cases},
\end{equation}
where
\begin{equation*}
\label{eq:BB}
u_{1}=
\begin{cases}
\left[0,U_\mathrm{p}\right]^\top, & \! e_\mathrm{B}(t)\leq 0\\
\left[U_\mathrm{a},0\right]^\top, & \! e_\mathrm{B}(t)>0
\end{cases},
\; \; u_{2}=
\begin{cases}
\left[U_\mathrm{a}, 0\right]^\top, & \! e_\mathrm{A}(t)\leq 0\\
\left[0,U_\mathrm{p}\right]^\top, & \! e_\mathrm{A}(t) > 0
\end{cases}.
\end{equation*}
%
%
%
%
\subsection{PI Controller}
The PI control inputs to be implemented via the Dial-a-Wave system are chosen as:
\begin{equation}
\label{eq:PIctrl}
\begin{split}
u_\mathrm{a}(t)  = \, & k_\mathrm{P,a}e_\mathrm{B}(t)+k_\mathrm{I,a}\int_{0}^{t}e_\mathrm{B}(t)dt\\
& -\left(k_\mathrm{P,p}e_\mathrm{A}(t)+k_\mathrm{I,p}\int_{0}^{t}e_\mathrm{A}(t)dt\right)
\end{split}
\end{equation}
with $k_\mathrm{P,p}$, $k_\mathrm{P,a}$,  $k_\mathrm{I,a} $ and $ k_\mathrm{I,p} $ being the control gains, and $u_\mathrm{p}(t)$ given by \eqref{eq:DAW}. 

Moreover, to improve the performance and guarantee that the control signals do not exceed their admissible values, the PI controller is complemented by a dynamic saturation defined as:
\begin{equation}\label{eq:dyn_sat}
\begin{cases}
	u_\mathrm{a}\in\left[0,50\right], & \text{if }  |e_\mathrm{B}|<|e_\mathrm{A}|\\
	u_\mathrm{a}\in\left[0,100\right], & \text{otherwise}
\end{cases}
\end{equation}
and an anti wind-up scheme.
\subsection{MPC Algorithm}
The last algorithm we considered is a Model Predictive Controller (MPC) \cite{mayne2000constrained}. 
Given the state of the cell population, say $\textbf{x}=[x_1, \dots, x_{N(t)}]^\top\in\mathbb{R}^{2N(t)}$, where $x_i=[LacI^i, TetR^i]^\top$ is the state of the $i$-th cell, we compute the optimal control input over the time interval $\left[t, t+T_{p}\right]$ which minimizes the cost function
\begin{equation}\label{eq:cost_fcn}
    J(\textbf{x},r,u,t)=\int_{0}^{Tp}\left(\alpha\|e_\mathrm{B}(t)\|+(1-\alpha)\|e_\mathrm{A}(t)\|\right)dt
    \end{equation}
with $T_p$ being the controller prediction time and $\alpha\in(0,1)$ a constant design parameter.
We then apply the computed optimal control input over the interval $\left[ t,t+T_c\right)$ where $T_c<T_p$ is a control time to be selected during the implementation.

To reduce the computational burden, in our implementation we evaluated the cost function $ J(\textbf{x},u) $ using only a representative subset of cells chosen as a sample of the entire population. 
This subset of cells is chosen such that the reduced ratios (i.e. the ratios $r_\mathrm{A}$ and $r_\mathrm{B}$ computed on the subset) is as close as possible to the ratios of the entire population. 
Also, a genetic algorithm taken from \cite{GA} (without the mutation phase) was used to find the optimal control sequence.
%
%
%
\section{In-silico validation and comparisons}
\label{sec:simulations}
We tested all the designed control laws in-silico. First, we performed a batch of simulations in Matlab assuming a constant population of 30 cells, then we performed more accurate simulations using an agent-based simulator specifically designed for bacterial populations called BSim \cite{BSim}.

In all simulations we consider a desired ratio $ r=0.6 $, with initial conditions taken in a neighborhood of the saddle point between \textbf{A} and \textbf{B}, specifically we picked the initial conditions for each cell with a random uniform distribution in the intervals: $ {mRNA_\mathrm{LacI,0}^{i}}\in \left[3,6\right] $, $ {mRNA_\mathrm{TetR,0}^{i}}\in \left[3,6\right] $, $ {LacI}_{0}^{i}\in \left[150,300\right] $, ${TetR}_{0}^{i}\in \left[200,400\right]$. 
To mimic the real experimental constraints, the state of the population is sampled every $T_s = 5\,\mathrm{min}$ and the control inputs are evaluated every $T_c = 15\, \mathrm{min}$.
%
%
%
%
\subsection{Numerical simulations in Matlab}
\begin{figure*}[t]
	\centering
	\subfloat[Bang-Bang controller: error signals] {\includegraphics[width=0.3\linewidth]{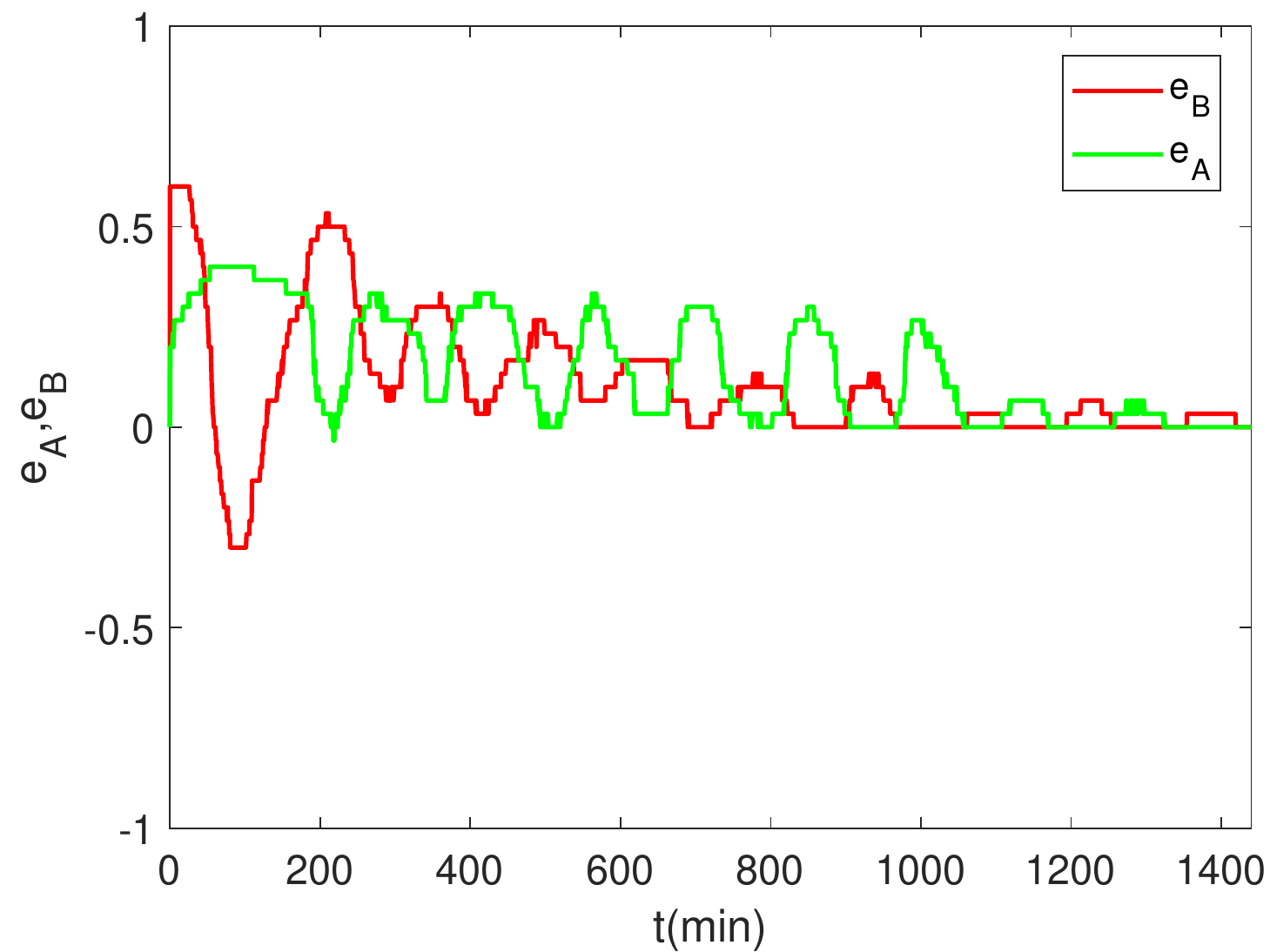}%
		\label{fig_first_case_nQS}}
	\hfil
	\setcounter{subfigure}{2}
	\subfloat[PI controller: error signals] {\includegraphics[width=0.3\linewidth]{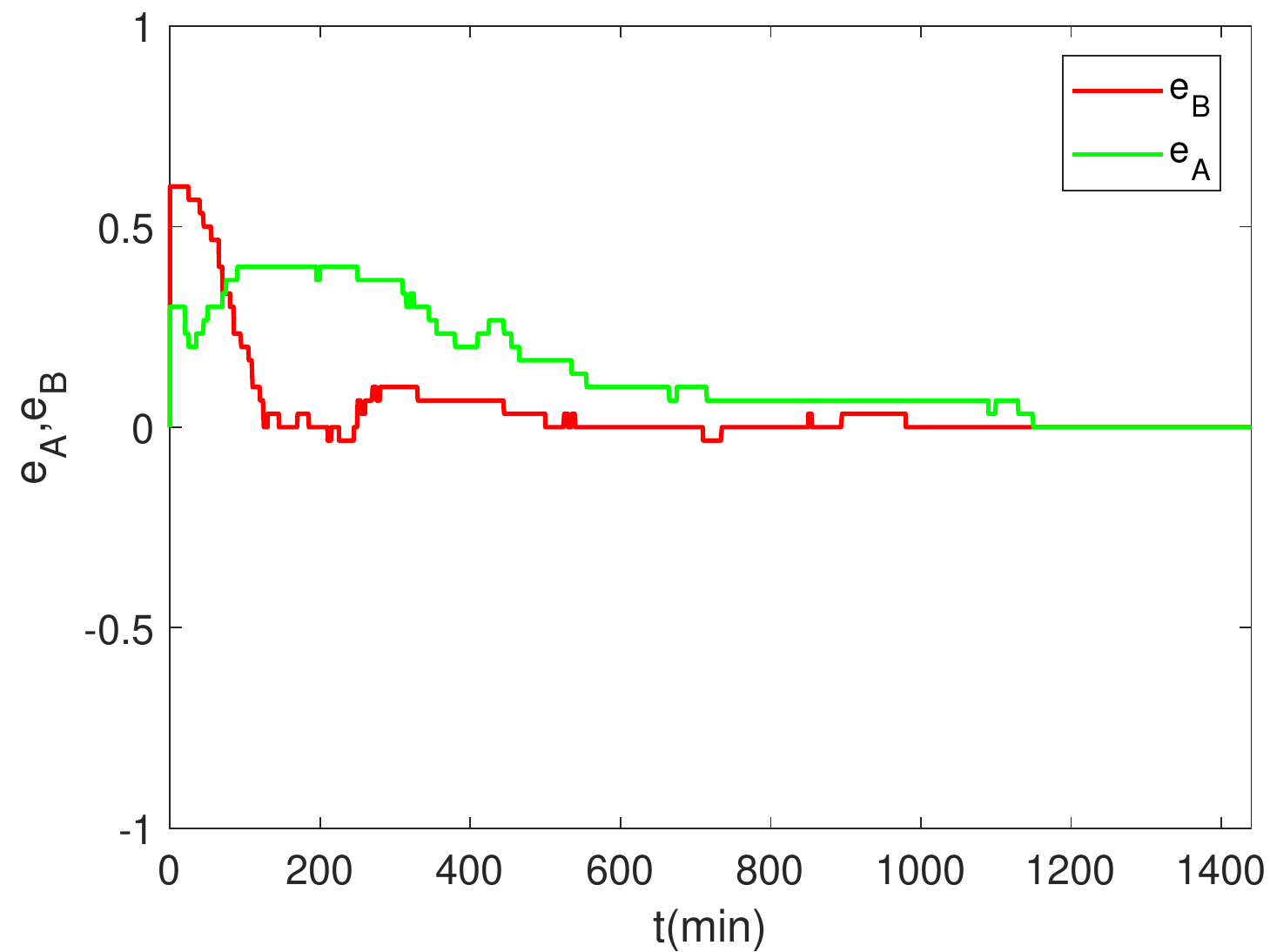}%
		\label{fig_second_case_nQS}}
	\hfil
	\setcounter{subfigure}{4}
	\subfloat[MPC: error signals] {\includegraphics[width=0.3\linewidth]{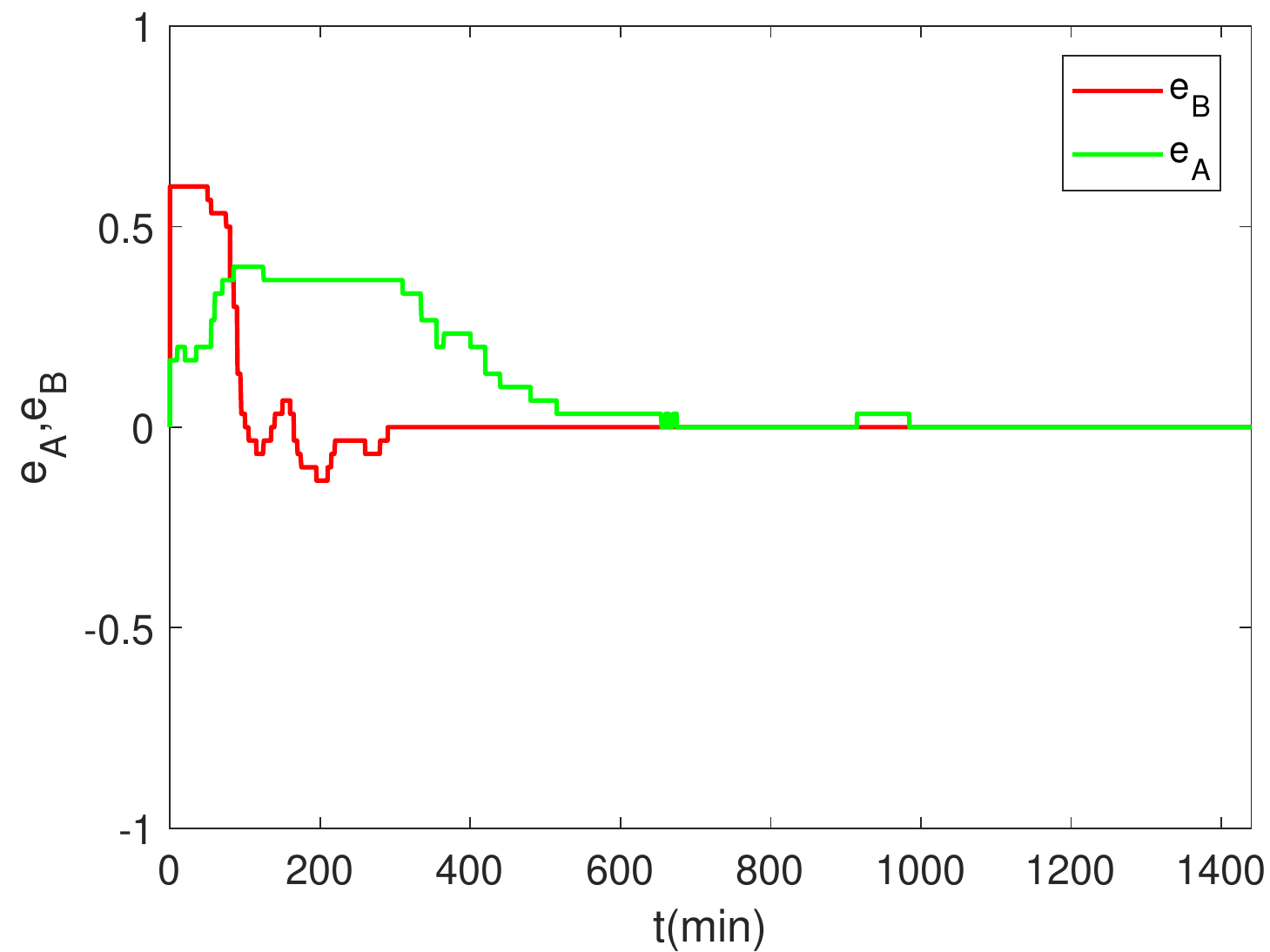}%
		\label{fig_third_case_nQS}}\\
	\setcounter{subfigure}{1}
	\subfloat[Bang-Bang controller: control inputs] {\includegraphics[width=0.3\linewidth]{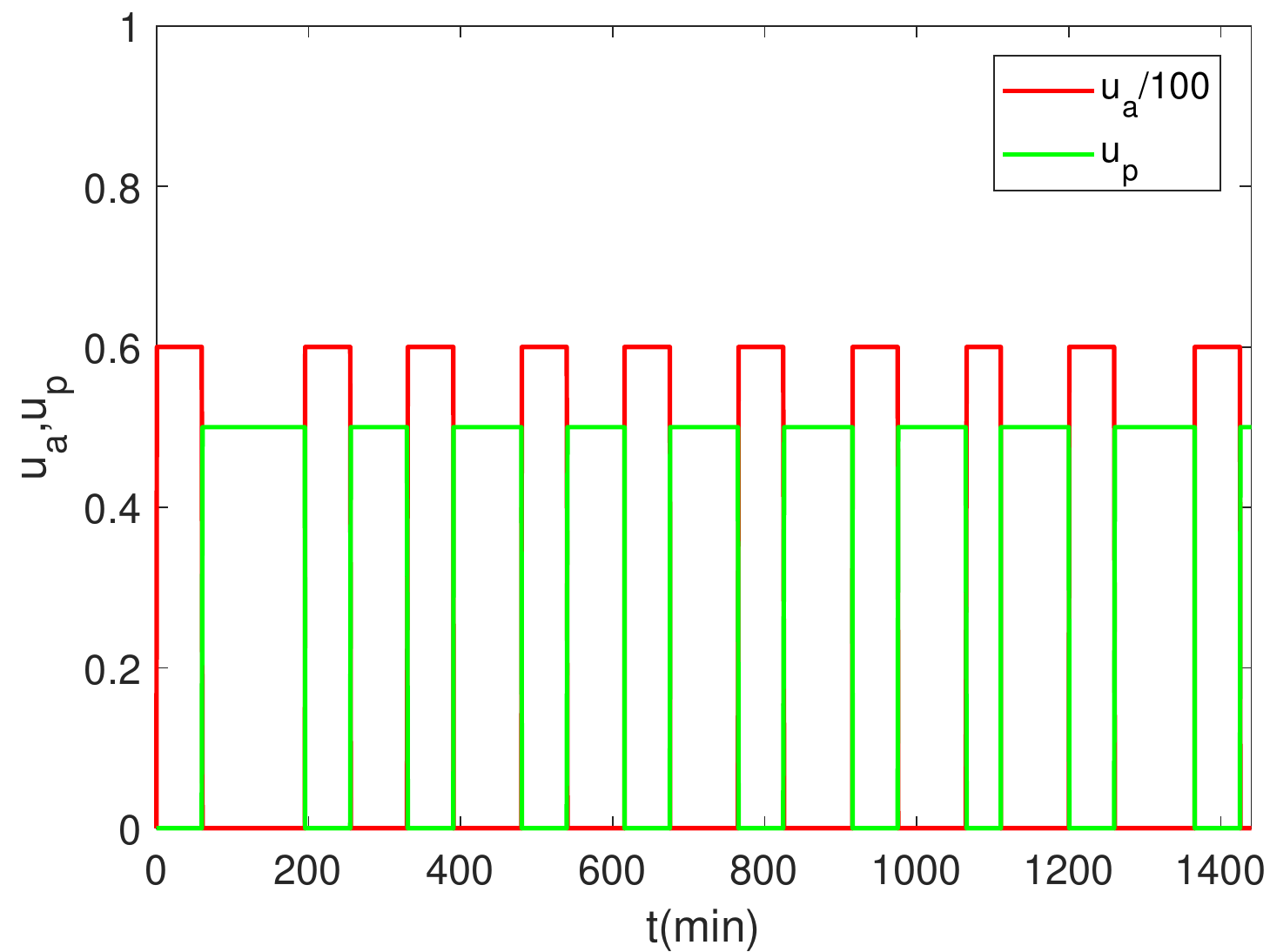}%
		\label{fig_first_case_2_nQS}}
	\hfil
	\setcounter{subfigure}{3}
	\subfloat[PI controller: control inputs] {\includegraphics[width=0.3\linewidth]{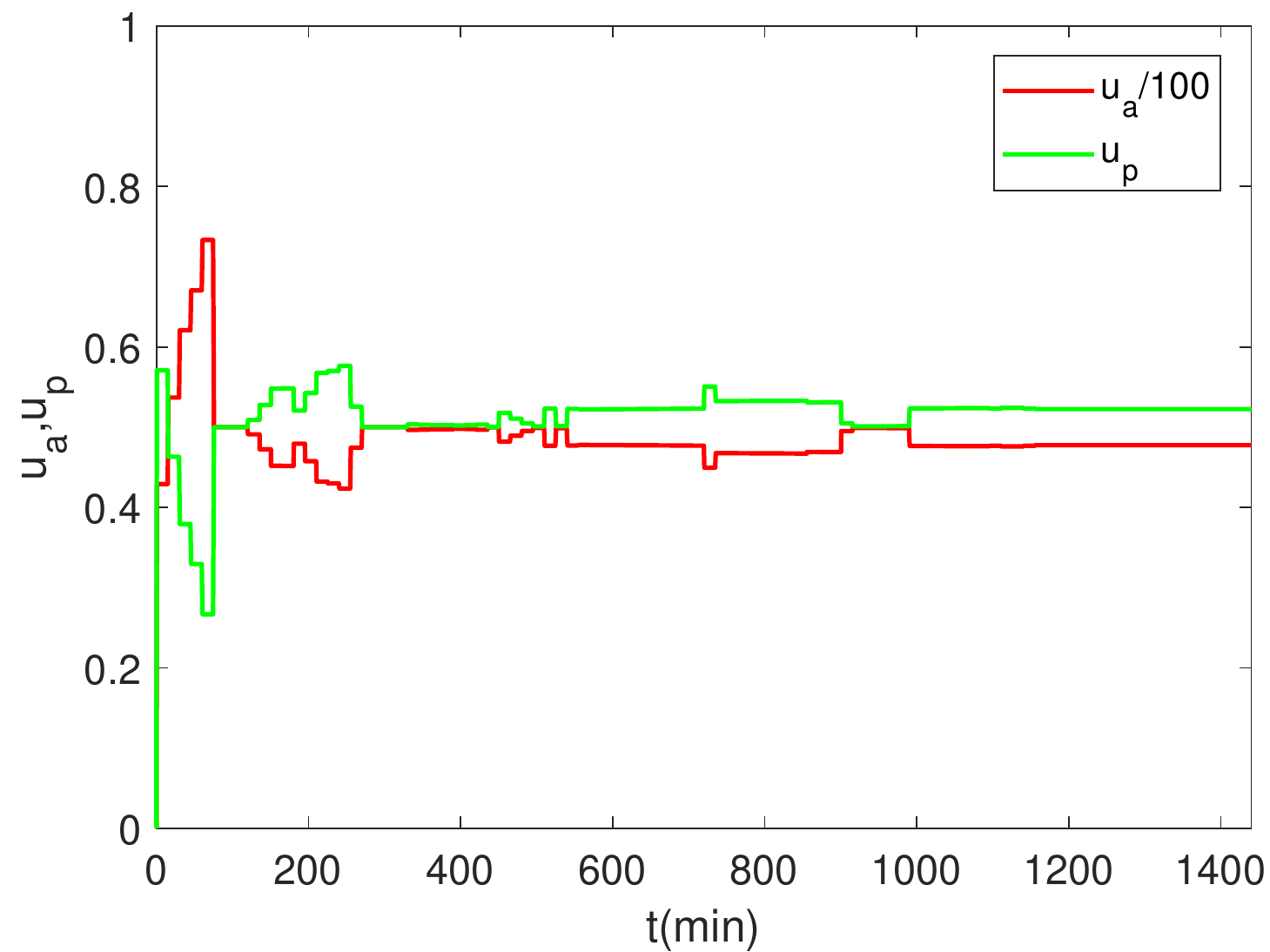}%
		\label{fig_second_case_2_nQS}}
	\hfil
	\setcounter{subfigure}{5}
	\subfloat[MPC: control inputs] {\includegraphics[width=0.3\linewidth]{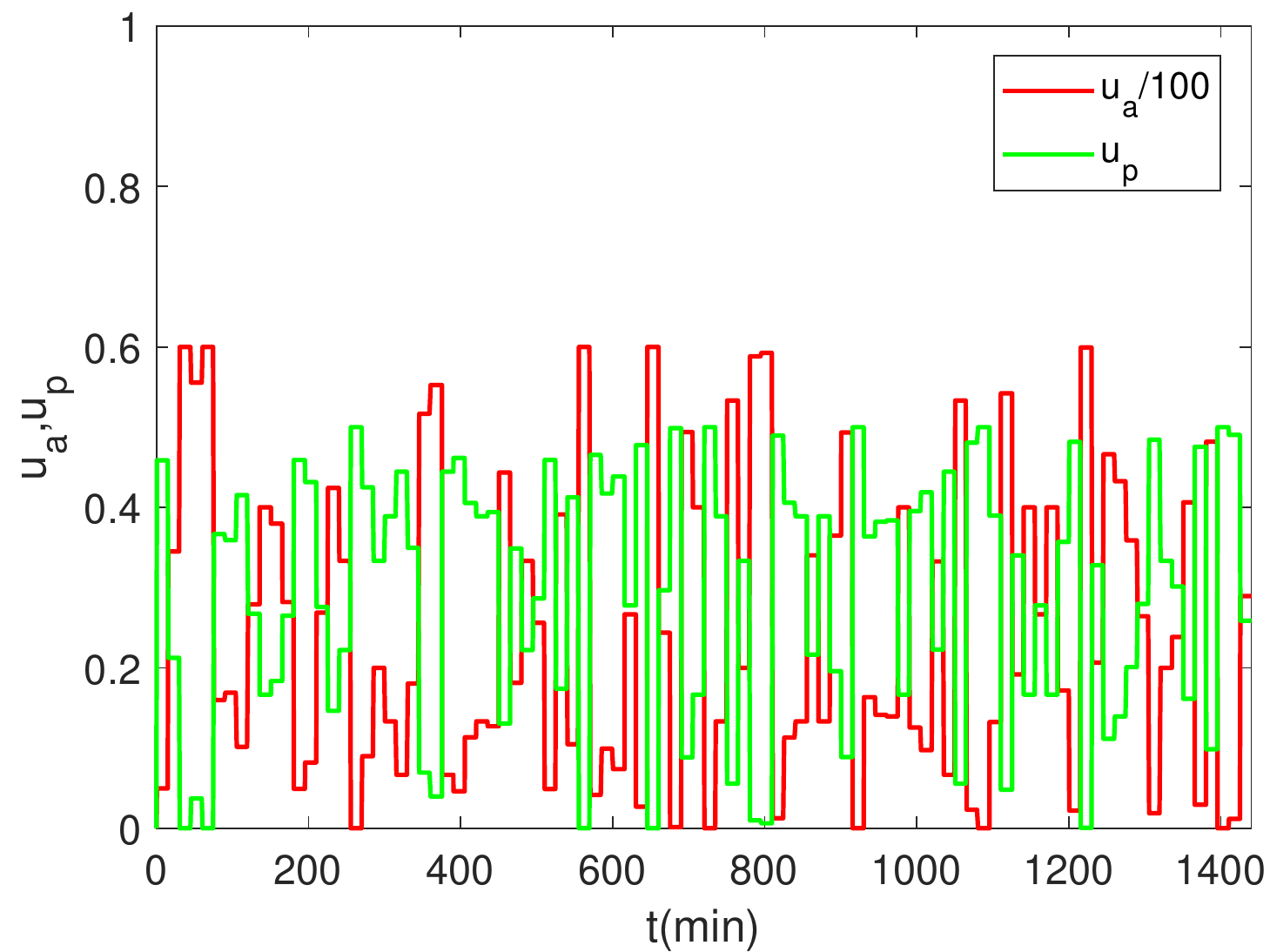}%
		\label{fig_third_case_2_nQS}}
	\caption{Evolution in time of the error signals $e_\mathrm{A}(t)$ and $e_\mathrm{B}(t)$ and of the control inputs $u_\mathrm{a}(t)$ and $u_\mathrm{p}(t)$ for a consortium of cells embedding an inducible toggle switch.
	}
	\label{fig:nQS}
\end{figure*}

For the Bang-Bang controller we empirically set the control amplitudes to $U_\mathrm{a}=60$ and $U_\mathrm{p}=0.5$. With these values we obtained the evolution of errors and control inputs shown in panels (a) and (b) of Figure \ref{fig:nQS}. We notice that the Bang-Bang controller achieves good performance with a settling time of about 1300 min. 

For the PI controller, the control input amplitudes and control gains were empirically selected as $U_\mathrm{a}=100$, $U_\mathrm{p}=1$, $k_\mathrm{P,a}=66.67$, $k_\mathrm{P,p}=2.25$, $k_\mathrm{I,a}=1.2$ and $k_\mathrm{I,p}=0.006 $ obtaining the results portrayed in panels (c), (d) of Figure \ref{fig:nQS}. In this case we observe a settling time which is half the one obtained with the Bang Bang controller, together with lower error values at steady state.

Finally, for the MPC controller, we chose $[U_\mathrm{a}, U_\mathrm{p}] = [60, 0.5]$, $T_{p}=75\,\mathrm{min}$, $ T_{c}=T_{s}=15\,\mathrm{min} $, $\alpha=0.6$. The genetic algorithm's parameters were set to $N_\mathrm{p}=20$, $ M_{\max}=10$, where $N_p$ is the length of the control sequence generated at each step and $M_{\max}$ the number of generations being considered.

The resulting errors and the control inputs are shown in panels (e) and (f) of Figure \ref{fig:nQS}, which confirm the MPC as the best strategy with a settling time almost $40 \%$ shorter than the one observed with the PI.
\subsection{Agent-based simulations in BSim}
To provide a more realistic validation of our control strategies, we used BSim \cite{BSim}, a realistic agent-based simulator of bacterial populations, which considers also cells reproduction, spatial distribution and geometry, the diffusion of the chemicals in the environment and, more importantly, the flush-out of cells from the chamber.
To run the required stochastic simulations, we extended BSim  with an Euler-Maruyama solver \cite{Stoc_sim}.

As a reference for the geometry of the microfluidic device we used a scaled version of the one described in \cite{danino2010synchronized} with dimensions $ 13.3 \mu \mathrm{m}\times 16.6 \mu \mathrm{m} \times 1 \mu \mathrm{m} $ which can contain a population of about $50$ cells.

\begin{figure*}[t]
	\centering
	\subfloat[Bacteria at t=0min] {\includegraphics[width=0.22\linewidth]{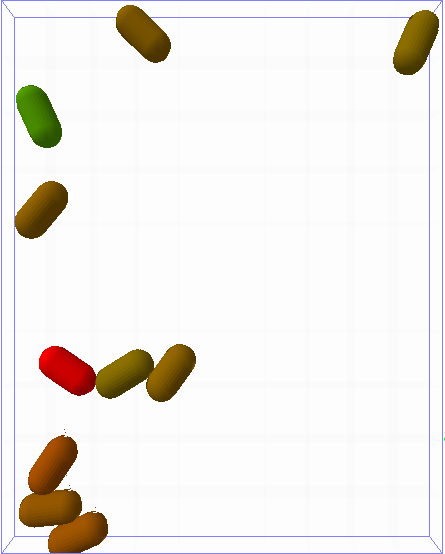}%
		\label{fig_BSim_snap1}}
	\hfil
	\subfloat[Bacteria at t=500min] {\includegraphics[width=0.22\linewidth]{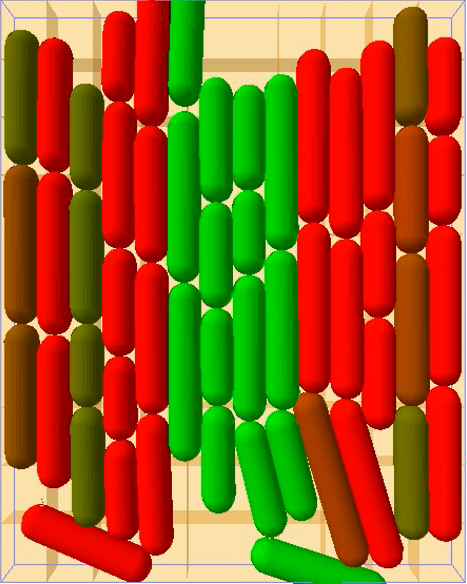}%
		\label{fig_BSim_snap2}}
		\hfil
	\subfloat[Bacteria at t=1440min] {\includegraphics[width=0.22\linewidth]{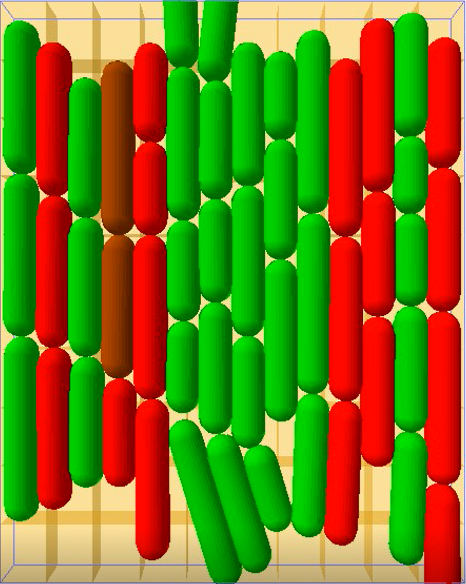}%

	\label{fig_BSim_snap3}}
    
    \subfloat[Control input evolution] {\includegraphics[width=0.9\linewidth]{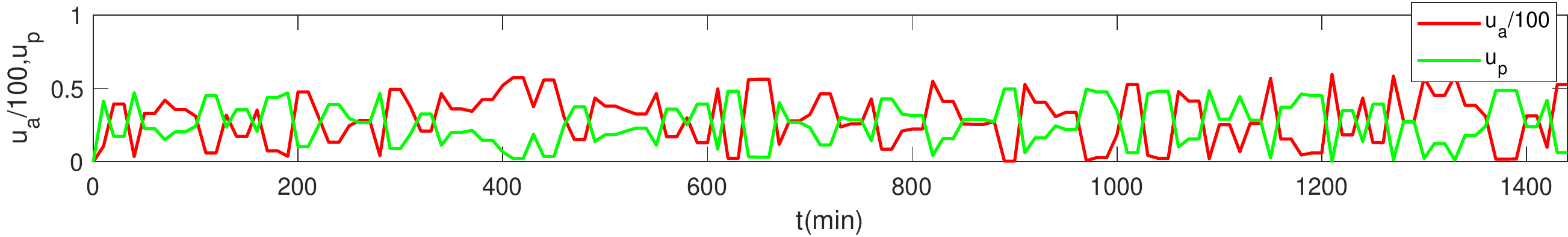}%
    
    \label{fig_BSim_u}}

	\caption{Snapshots of a BSim simulation performed using the MPC algorithm. Here, red cells belongs to $\mathcal{A}_t$ while green denotes cells belonging to $\mathcal{B}_t$. Bacteria with an intermediate coloration belong to $\mathcal{C}_t$. The corresponding control input are shown in the bottom panel. }
	\label{fig:BSIM_snap}
\end{figure*}

We tested in BSim all the proposed controllers and snapshots of a typical BSim simulation is shown in Figure \ref{fig:BSIM_snap} where cells fully expressing one of the two repressor genes are depicted either in red or green.

\begin{figure*}[t]
	\centering
	\subfloat[Bang-Bang controller] {\includegraphics[width=0.3\linewidth]{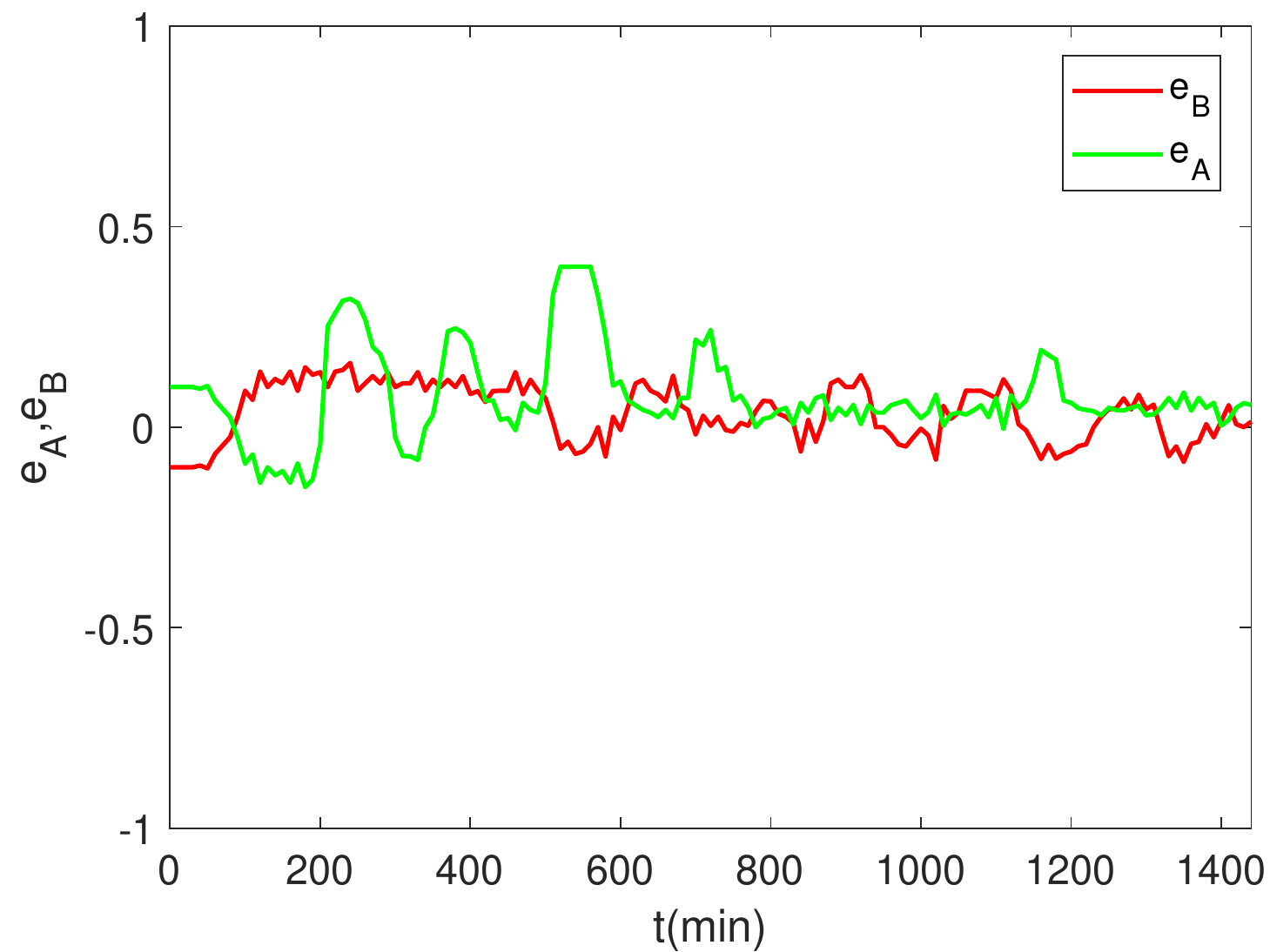}%
		\label{fig_BSimBB_nQS}}
	\hfil
	\subfloat[PI controller] {\includegraphics[width=0.3\linewidth]{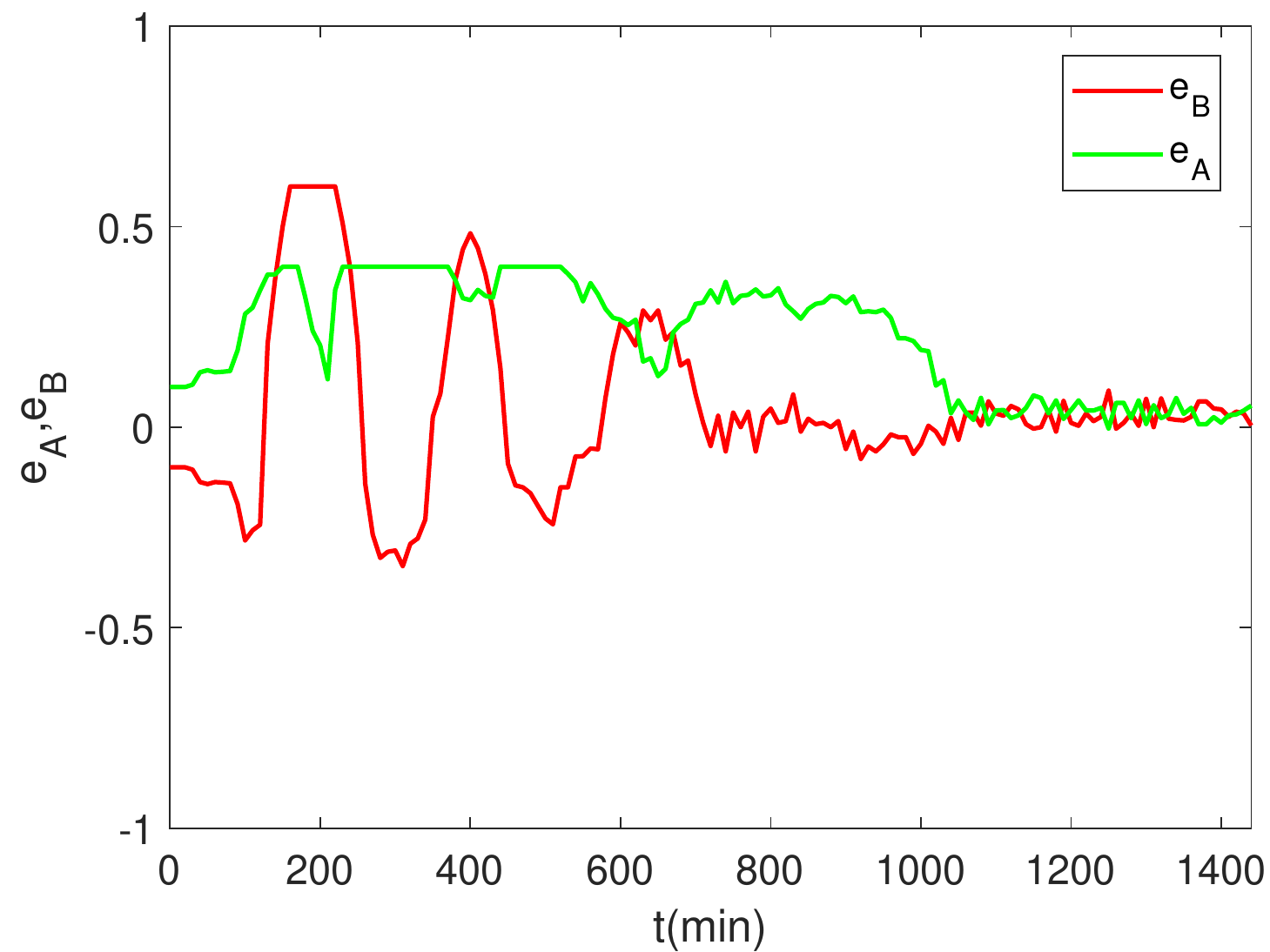}%
		\label{fig_BSimPI_nQS}}
		\hfil
	\subfloat[MPC] {\includegraphics[width=0.3\linewidth]{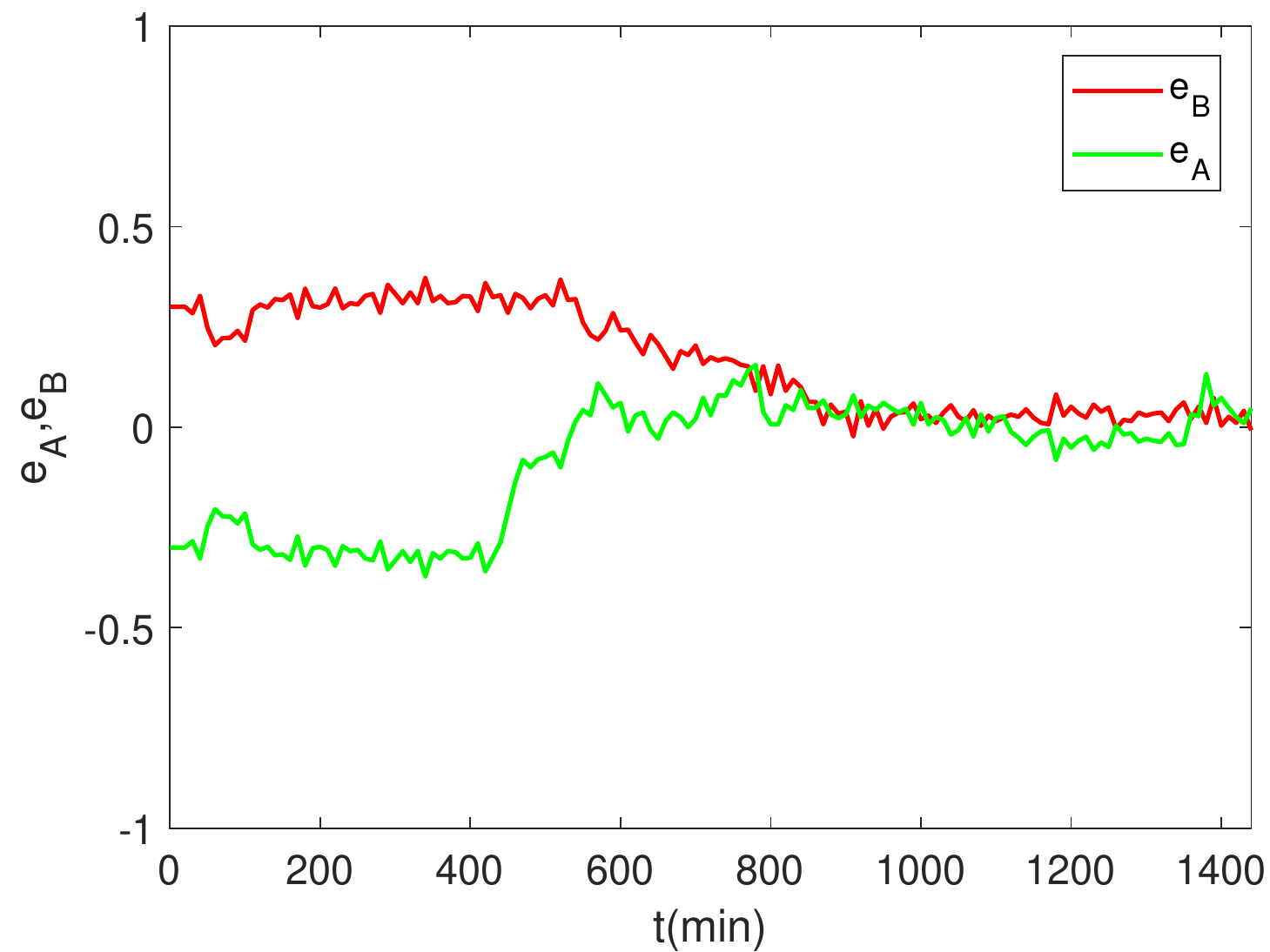}%
		\label{fig_BSimMPC_nQS}}
	\caption{BSim simulations: evolution in time of the error signals $e_\mathrm{A}(t)$ and $e_\mathrm{B}(t)$  for a consortium of cells embedding an inducible toggle switch.}
	\label{fig:BSIM_nQS}
\end{figure*}

The errors obtained via BSim \emph{in-silico} experiments are reported in Figure \ref{fig:BSIM_nQS}. It can be noticed that the fluctuations of the error signals are higher than in the previous Matlab simulations essentially due to cell growth and splitting, and the flush-out of the cells from the chamber. However, the average error evolution is qualitatively the same, confirming the good performance of the controllers.
\subsection{Performance Comparison}
To compare the performance of the controllers, we considered the following performance indices evaluated averaging over $M$ simulation trials: 

    (i) the average value of the error norm over the total simulation time $T_\mathrm{sim}$,
    $
    \bar{e} = \frac{1}{M} \sum_{j=1}^M \left( \frac{1}{T_\mathrm{sim}} \int_0^{T_\mathrm{sim}} \lVert e_j(t) \rVert dt \right),
    $
    and (ii) over the last $180\,\mathrm{min}$,
    $
    \bar{e}_\mathrm{f} = \frac{1}{M} \sum_{j=1}^M \left( \frac{1}{180} \int_{T_\mathrm{sim}-180}^{T_\mathrm{sim}} \lVert e_j(t) \rVert dt \right),
    $
    where for the $j$-th trial $e_j(t) = [e_\mathrm{A}^j(t), \, e_\mathrm{B}^j(t)]^\top$,

    and (iii) the average settling time at $15\%$ of the error, $\bar{t}_\mathrm{s}$.

In Table \ref{tab:nearRatio} we report the values of the previous indices considering $M=30$ for the simulations in  Matlab and $M=1$  for those in BSim. It can be observed that the MPC algorithm guarantees excellent performance both in terms of settling time and steady-state error norm. Therefore, it is the best candidate for the {\em in-vivo} implementation that will be the next stage of our ongoing research. 
Nevertheless, the Bang-Bang controller offers a good compromise between ease of implementation and  performance.
\begin{table} [!t]
	\centering
	\begin{tabular}{|c|c|c|c|c|c|c|}
		\hline 
		Controller & $\bar{e}$ & $\bar{e}_\mathrm{f}$ & $\bar{t}_{s}$ \,(min) \\ 
		\hline 
		Bang-Bang & 0.20 (0.13) & 0.07 (0.06) & 1077 (1185)\\
		\hline 
		PI & 0.17 (0.28) & 0.02 (0.05) & 563 (1020)\\
		\hline
		MPC & 0.13 (0.22) & 0.02 (0.05) & 329 (820)\\
		\hline
	\end{tabular} 
	\caption{Performance indices of the proposed controllers evaluated using Matlab (BSim, respectively) simulations.
	}
	\label{tab:nearRatio}
\end{table}
%
%
%
%
%
\section{Conclusions}
We considered the ratiometric control problem in a mono-strain microbial consortium made of bacteria embedding a bistable toggle switch. We demonstrated that, by varying global inputs to all the cells, it is possible to control the ratio between those stabilizing onto one equilibrium and those on the other. 
Namely, we presented three control strategies to regulate the cells in the consortium to the desired ratio.
The control design took into account the constraints of a possible experimental microfluidic implementation. We tested performance of the controllers {in-silico} by numerical and realistic agent-based simulations. In both cases, it emerged that the MPC algorithm guarantees excellent performance in terms of both the settling time and the steady-state error values.
Future work will be aimed at validating \emph{in-vivo} the proposed controllers and exploiting them for multicellular feedback control schemes such as the one described in \cite{fiore2016silico}.

\footnotesize
\section*{ACKNOWLEDGMENT}
The authors wish to acknowledge support from the research project COSY-BIO (Control Engineering of Biological Systems for Reliable Synthetic Biology Applications) funded by the European Union's Horizon 2020 research and innovation programme under grant agreement No 766840.

\bibliographystyle{IEEEtran} 
\bibliography{refs} 

\begin{thebibliography}{10}
\providecommand{\url}[1]{#1}
\csname url@samestyle\endcsname
\providecommand{\newblock}{\relax}
\providecommand{\bibinfo}[2]{#2}
\providecommand{\BIBentrySTDinterwordspacing}{\spaceskip=0pt\relax}
\providecommand{\BIBentryALTinterwordstretchfactor}{4}
\providecommand{\BIBentryALTinterwordspacing}{\spaceskip=\fontdimen2\font plus
\BIBentryALTinterwordstretchfactor\fontdimen3\font minus
  \fontdimen4\font\relax}
\providecommand{\BIBforeignlanguage}[2]{{%
\expandafter\ifx\csname l@#1\endcsname\relax
\typeout{** WARNING: IEEEtran.bst: No hyphenation pattern has been}%
\typeout{** loaded for the language `#1'. Using the pattern for}%
\typeout{** the default language instead.}%
\else
\language=\csname l@#1\endcsname
\fi
#2}}
\providecommand{\BIBdecl}{\relax}
\BIBdecl

\bibitem{del2018future}
D.~Del~Vecchio, Y.~Qian, R.~M. Murray, and E.~D. Sontag, ``Future systems and
  control research in synthetic biology,'' \emph{{Annual Reviews in Control}},
  vol.~45, pp. 5--17, 2018.

\bibitem{bittihn2018rational}
P.~Bittihn, M.~O. Din, L.~S. Tsimring, and J.~Hasty, ``Rational engineering of
  synthetic microbial systems: from single cells to consortia,'' \emph{{Current
  Opinion in Microbiology}}, vol.~45, pp. 92--99, 2018.

\bibitem{balagadde2008synthetic}
F.~K. Balagadd{\'e}, H.~Song, J.~Ozaki, C.~H. Collins, M.~Barnet, F.~H. Arnold,
  S.~R. Quake, and L.~You, ``{A synthetic Escherichia coli predator--prey
  ecosystem},'' \emph{{Molecular Systems Biology}}, vol.~4, no.~1, p. 187,
  2008.

\bibitem{chen2015emergent}
Y.~Chen, J.~K. Kim, A.~J. Hirning, K.~Josi{\'c}, and M.~R. Bennett, ``Emergent
  genetic oscillations in a synthetic microbial consortium,'' \emph{Science},
  vol. 349, no. 6251, pp. 986--989, 2015.

\bibitem{sadeghpour2017bistability}
M.~Sadeghpour, A.~Veliz-Cuba, G.~Orosz, K.~Josi{\'c}, and M.~R. Bennett,
  ``Bistability and oscillations in co-repressive synthetic microbial
  consortia,'' \emph{{Quantitative Biology}}, vol.~5, no.~1, pp. 55--66, 2017.

\bibitem{fiore2016silico}
G.~Fiore, A.~Matyjaszkiewicz, F.~Annunziata, C.~Grierson, N.~J. Savery,
  L.~Marucci, and M.~di~Bernardo, ``In-silico analysis and implementation of a
  multicellular feedback control strategy in a synthetic bacterial
  consortium,'' \emph{{ACS Synthetic Biology}}, vol.~6, no.~3, pp. 507--517,
  2016.

\bibitem{igem}
\BIBentryALTinterwordspacing
{Imperial College London}. {2016 iGEM - Ecolibrium project}. [Online].
  Available: \url{http://2016.igem.org/Team:Imperial_College}
\BIBentrySTDinterwordspacing

\bibitem{ren2017population}
X.~Ren, A.-A. Baetica, A.~Swaminathan, and R.~M. Murray, ``Population
  regulation in microbial consortia using dual feedback control,'' \emph{{Proc.
  of the 56th IEEE Conference on Decision and Control (CDC)}}, pp. 5341--5347,
  2017.

\bibitem{Gar}
T.~S. Gardner, C.~R. Cantor, and J.~J. Collins, ``{Construction of a genetic
  toggle switch in Escherichia coli},'' \emph{Nature}, vol. 403, no. 6767, p.
  339, 2000.

\bibitem{BSim}
A.~Matyjaszkiewicz, G.~Fiore, F.~Annunziata, C.~S. Grierson, N.~J. Savery,
  L.~Marucci, and M.~di~Bernardo, ``Bsim 2.0: an advanced agent-based cell
  simulator,'' \emph{{ACS Synthetic Biology}}, vol.~6, no.~10, pp. 1969--1972,
  2017.

\bibitem{lugagne}
J.-B. Lugagne, S.~S. Carrillo, M.~Kirch, A.~K{\"o}hler, G.~Batt, and P.~Hersen,
  ``Balancing a genetic toggle switch by real-time feedback control and
  periodic forcing,'' \emph{Nature Communications}, vol.~8, no.~1, p. 1671,
  2017.

\bibitem{fiore2019analysis}
D.~Fiore, A.~Guarino, and M.~di~Bernardo, ``Analysis and control of genetic
  toggle switches subject to periodic multi-input stimulation,'' \emph{IEEE
  Control Systems Letters}, vol.~3, no.~2, pp. 278--283, 2019.

\bibitem{guarino2018silico}
A.~Guarino, D.~Fiore, and M.~di~Bernardo, ``In-silico feedback control of a
  {MIMO} synthetic toggle switch via pulse-width modulation,'' in \emph{Proc.
  of the 2019 European Control Conference (ECC)}, 2019.

\bibitem{lakatos2017stochastic}
E.~Lakatos, ``Stochastic analysis and control methods for molecular cell
  biology,'' Ph.D. dissertation, {Imperial College London}, 2017.

\bibitem{CME}
D.~T. Gillespie, ``A rigorous derivation of the chemical master equation,''
  \emph{{Physica A: Statistical Mechanics and its Applications}}, vol. 188, no.
  1-3, pp. 404--425, 1992.

\bibitem{CLE}
------, ``{The chemical Langevin equation},'' \emph{{The Journal of Chemical
  Physics}}, vol. 113, no.~1, pp. 297--306, 2000.

\bibitem{menolascina2014vivo}
F.~Menolascina, G.~Fiore, E.~Orabona, L.~De~Stefano, M.~Ferry, J.~Hasty,
  M.~di~Bernardo, and D.~di~Bernardo, ``In-vivo real-time control of protein
  expression from endogenous and synthetic gene networks,'' \emph{{PLoS
  Computational Biology}}, vol.~10, no.~5, p. e1003625, 2014.

\bibitem{Perrino2015}
G.~Fiore, G.~Perrino, M.~di~Bernardo, and D.~di~Bernardo, ``In vivo real-time
  control of gene expression: a comparative analysis of feedback control
  strategies in yeast,'' \emph{{ACS Synthetic Biology}}, vol.~5, no.~2, pp.
  154--162, 2015.

\bibitem{ferry2011microfluidics}
M.~S. Ferry, I.~A. Razinkov, and J.~Hasty, ``Microfluidics for synthetic
  biology: from design to execution,'' in \emph{Methods in Enzymology}.\hskip
  1em plus 0.5em minus 0.4em\relax Elsevier, 2011, vol. 497, pp. 295--372.

\bibitem{mayne2000constrained}
D.~Q. Mayne, J.~B. Rawlings, C.~V. Rao, and P.~O. Scokaert, ``Constrained model
  predictive control: Stability and optimality,'' \emph{Automatica}, vol.~36,
  no.~6, pp. 789--814, 2000.

\bibitem{GA}
D.~E. Goldberg and J.~H. Holland, ``Genetic algorithms and machine learning,''
  \emph{Machine learning}, vol.~3, no.~2, pp. 95--99, 1988.

\bibitem{Stoc_sim}
D.~J. Higham, ``An algorithmic introduction to numerical simulation of
  stochastic differential equations,'' \emph{SIAM Review}, vol.~43, no.~3, pp.
  525--546, 2001.

\bibitem{danino2010synchronized}
T.~Danino, O.~Mondrag{\'o}n-Palomino, L.~Tsimring, and J.~Hasty, ``A
  synchronized quorum of genetic clocks,'' \emph{Nature}, vol. 463, no. 7279,
  p. 326, 2010.

\end{thebibliography}

\end{document}